\documentclass[fleqn,usenatbib]{mnras}

\usepackage{newtxtext,newtxmath}
\usepackage[T1]{fontenc}
\usepackage{ae,aecompl}
\usepackage{soul}

\usepackage{graphicx}	% Including figure files
\usepackage{amsmath}	% Advanced maths commands
\usepackage{amssymb}	% Extra maths symbols
\usepackage{multirow} 	% Tables - multiple rows
\usepackage{lscape} 	% Tables - large tables need landscape orientation

\usepackage{chngcntr} 
\usepackage{chngcntr}
\counterwithout*{figure}{section}
\counterwithout*{table}{section}
\usepackage{xcolor}

\title[Identifying offset--AGN using gravitational lensing]{Constraining VLBI$-$optical offsets in high redshift galaxies using strong gravitational lensing}

\author[C.~Spingola \& A.~Barnacka]{Cristiana Spingola,$^{1,2}$\thanks{E-mail: spingola@ira.inaf.it}
and Anna Barnacka$^{3,4}$\thanks{E-mail: abarnacka@cfa.harvard.edu}
\\
$^{1}$ INAF $-$ Istituto di Radioastronomia, Via Gobetti 101, I$-$40129, Bologna, Italy \\
$^{2}$ Dipartimento di Fisica e Astronomia, Universit\`a degli Studi di Bologna, Via Gobetti 93/2, I$-$40129 Bologna, Italy \\
$^{3}$ Harvard$-$Smithsonian Center for Astrophysics, 60 Garden St, MS$-$20, Cambridge, MA 02138, USA \\
$^{4}$ Astronomical Observatory, Jagiellonian University, Cracow, Poland\\
}

\date{Accepted 2020 March 24. Received 2020 March 17; in original form 2020 February 6.
}

\pubyear{2020}

\begin{document}
\label{firstpage}
\pagerange{\pageref{firstpage}--\pageref{lastpage}}
\maketitle

% Abstract of the paper, currently 237 words
\begin{abstract}
We present a multi-wavelength analysis of two highly magnified strong gravitationally lensed galaxies, CLASS~B0712+472 and CLASS~B1608+656, at redshifts $1.34$ and $1.394$, respectively, using new VLBI and archival {\it HST} observations. We reconstruct the positions of the radio and optical emissions with their uncertainties using Monte Carlo sampling. We find that in CLASS~B0712+472 the optical and radio emissions are co-spatial within $2\pm5$~mas ($17\pm 42$~pc at redshift of 1.34). But, in CLASS~B1608+656, we reconstruct an optical--radio offset of $25\pm16$~mas ($214\pm137$~pc at redshift of 1.394), the smallest offset measured for an AGN at such high redshift. The spectral features indicate that CLASS~B1608+656 is a post-merger galaxy, which, in combination with the optical--VLBI offset reported here, makes CLASS~B1608+656 a promising candidate for a high-$z$ offset--AGN. Furthermore, the milliarcsecond angular resolution of the VLBI observations combined with the precise lens models allow us to spatially locate the radio emission at $0.05$~mas precision ($0.4$~pc) in CLASS~B0712+472, and $0.009$~mas precision ($0.08$~pc) in CLASS~B1608+656. The search for optical--radio offsets in high redshift galaxies will be eased by the upcoming synoptic all-sky surveys, including E-ELT and SKA, which are expected to find $\sim 10^5$ strongly lensed galaxies, opening an era of large strong lensing samples observed at high angular resolution.

\end{abstract}
% Select between one and six entries from the list of approved keywords.
% Don't make up new ones.

%%%%%%%%%%%%%%%%%%%%%%%%%%%%%%%%%%%%%%%%%%%%%%%%%%

%%%%%%%%%%%%%%%%% BODY OF PAPER %%%%%%%%%%%%%%%%%%

\section{Introduction}
\label{sec:introduction}

The centres of galaxies are powerful laboratories to test theoretical predictions of galaxy formation and evolution (\citealt{White1978}, for recent review see \citealt{Vogelsberger2019}). For example, observations and simulations indicate that properties of super-massive black holes (SMBHs) are highly correlated with central stellar nuclei, implying a strong co-evolution of SMBH and their host galaxies \citep[e.g.][]{Merritt2001, Volonteri2015}. 

SMBH are expected to grow via galaxy mergers and accretion of their surrounding material (stars and gas), which can trigger a phase of their evolution referred to as an active galactic nucleus (AGN, \citealt{Padovani2017}). Therefore, galaxy mergers are interesting targets for investigating the origins of AGN triggering. In the galaxy merger scenario, bright AGN should be preferentially found in interacting galaxies  and close SMBH pairs should be found at the late stages of the merger of galaxies \citep[e.g.][]{Comerford2014}. Furthermore, the efficiency of merger-triggered AGN should be higher for interacting galaxies with separation below 10~kpc \citep{Steinborn2016}. 

Pairs of SMBH separated by a few kpcs may be observed as \textsl{dual} AGN system when both SMBHs are actively accreting.  However, if only one SMBH is accreting and its X-ray/radio emissions is offset with respect to the optical emission of the host galaxy then such systems are identified as \textsl{offset--AGN} \citep{Comerford2014, Burke-Spolaor2018}.  The nature of offset--AGN is complex. Such systems can also consist of a single recoiling  SMBH (e.g. \citealt{Lena2014} and references therein). Recoiling SMBH after SMBH--SMBH mergers are a main target of the \textsl{Laser Interferometer Space Antenna} (LISA) for detecting gravitational waves at the sub-Hz regime \citep[e.g.][]{Volonteri2008, Blecha2016}. Therefore, identification of offset--AGN systems can provide valuable information on SMBH formation and evolution, as well as the triggering mechanisms of AGN. However, identifying multi-wavelength offsets at sub-kpc scales is severely limited by the angular resolution and the astrometry of current facilities. The studies are further limited for high redshift sources which could provide critical observations of galaxy evolution and abundance of offset--AGN \citep{Komossa2012, Orosz2013, Lena2014, Barrows2016, Skipper2018, Blecha2018, Rosas-Guevara2018}.

The magnification provided by gravitational lensing can ease the identification of high-$z$ offset--AGN. A gravitationally lensed system consists of a massive object (lens) that bends light rays from a background galaxy (source). As a result, multiple images of the background source can form (see \citealt{Treu2010} for a review).  Furthermore, ellipticity in gravitational potential of the lens can result in creation of complex caustic patterns. A caustic network is the locus in the source plane of points which would be magnified infinitely by the lens. Magnification and positions of lensed images change rapidly depending on the distance from the source to the caustics.  Such non-linear amplification in the position of lensed images and magnifications can be used to improve spatial resolution by orders of magnitude as proposed by \citet{Barnacka2017,Barnacka2018} (we refer to \citealt{Congdon2018} for an in-depth description of critical and caustic curves). The combination of gravitational lensing with high angular resolution observations can give access to the pc and sub-pc scales of distant galaxies, and can be used to reveal offset--AGN even at high redshift \citep[e.g.][]{Deane2013, Spingola2019c, Hartley2019}. 

In this paper, we demonstrate the resolving power of gravitational lensing to search for possible optical--radio offsets at mas-level for sources located at cosmological distances. We investigate the relative location of optical- and radio-emitting regions in two gravitationally lensed high redshift galaxies, CLASS~B0712+472 and CLASS~B1608+656, located close to the caustics and, therefore, highly magnified. 

This paper is structured as follows. In Section~\ref{sec:method}, we provide an overview of the method of using gravitational caustics of lensing galaxies to improve spatial resolution. In Section~\ref{sec:targets}, we introduce the targets CLASS B0712+472 and CLASS B1608+656. Section~\ref{sec:observations} provides a description of the data reduction process for both the optical and radio observations. In Section~\ref{sec:source_inversion}, we present the source inversion process and the statistical analysis of positional uncertainties.  We present the results in Section~\ref{sec:results}, and discuss their implication in Section~\ref{sec:discussion}. Finally, we summarise our results in Section~\ref{sec:conclusions}. 
Throughout this paper, we assume $H_0$ = 67.8 km s$^{-1}$ Mpc$^{-1}$, $\Omega_M$ = 0.31, and $\Omega_{\Lambda}$= 0.69 \citep{Planck2016}.

\section{Methodology}
\label{sec:method}
In this section, we summarise the steps that we perform to search for possible spatial offsets between the radio and optical emissions in gravitationally lensed AGN systems. To overcome the current technological limitations, \citet{Barnacka2017} proposed a method for spatially resolving the multi-wavelength emission in high-$z$ galaxies that takes advantage of the caustic curves, which act as non-linear amplifiers and, therefore, enhance the performance of telescopes (see also \citealt{Barnacka2018}). The caustics are the locations where the lensed images merge (or are created): if the source is located outside the inner caustics but inside the outer caustics of the lensing galaxy, then, typically, two lensed images of the same source are observed.  Otherwise, if the source is located inside the inner caustics, then four images of the same background source are created. 

The closer the source is to the caustics, the higher the magnification and amplification. Thus, if an AGN is located close to a caustic curve, even a small offset between the optical and radio emitting regions can result in a large difference in the positions of their lensed images.  For example, if the optical and radio emissions are at the same location, the position of the lensed images at radio and optical wavelengths coincides. However, if there is an offset between the radio and optical emission of, for instance, $1$~mas, then the difference in the position of the optical and radio lensed images can be offset by even $100$~mas \citep{Barnacka2017}.  Resolving offsets at the milliarcsecond-level between the optical and radio emissions is challenging, and, to date, can be performed mainly for local sources \citep[e.g.]{Plavin2019}. However, an offset of $100$~mas can be more easily resolved using even existing facilities.  The detection of such offsets in gravitationally lensed images would provide direct evidence for a physical spatial separation between the optical and radio emissions in the source plane. We proceed as follows to measure offsets between radio and optical emissions.

\noindent \textbf{Step 1.} Measuring the relative position of the lensed images at radio and optical wavelengths (Sec.~\ref{sec:VLBA_observations} and \ref{sec:hst_observations}).
 In order to align the images on the same coordinate frame (the radio VLBI), we perform a linear transformation that minimizes the difference between the lensed images positions as measured in the optical and the VLBI images. We found this alignment to be completely consistent with that obtained by using the WCS information within the optical positional uncertainties of the lensed images (given in Tables~\ref{tab:b0712_positions} and \ref{tab:b1608_positions}). Then, we measure the position of the lensed images in the {\it HST} and VLBI images aligned on the same reference frame.

\noindent \textbf{Step 2}. Backward ray-tracing the radio and optical emissions using a precise lens mass model. In this step we recover the source position  in each separate waveband (Sec.~\ref{sec:lensmodelling}). 

\noindent \textbf{Step 3}. Combining the uncertainties on the lens mass model parameters and observed positions of lensed images using a Monte Carlo sampling. With this approach, the error on the reconstructed source position at both optical and radio wavelengths takes into account both the uncertainties due to the observations and those of the lens mass model (Sec.~\ref{sec:uncertainty_source}). 

\noindent \textbf{Step 4}. Estimating the projected distance between the optical and radio emissions, and evaluating its significance (Sec.~\ref{sec:results}).

\section{Gravitationally lensed systems}
\label{sec:targets}

We select two flat-spectrum radio-loud gravitationally lensed systems with sources located close to the caustics. Selected systems, CLASS~B0712+472 and CLASS~B1608+656, are both quadruply-imaged in fold configuration. 
Therefore, they are ideal to use their properties as lensed systems for spatially locating the multi-wavelength emission \citep{Barnacka2017}. Below, we provide an overview of the properties of these two systems.

\subsection{CLASS~B0712+472} 
CLASS~B0712+472 was discovered as part of the Cosmic Lens All-Sky Survey (CLASS; \citealt{Jackson1998, Myers2003, Browne2003}). The system consists of a quadruply-imaged radio-loud AGN at redshift $z_s = 1.34$ and a lensing galaxy at $z_l = 0.41$ \citep{Fassnacht1998}. The lensing galaxy is part of a cluster of galaxies \citep{Fassnacht2002, Wilson2017}, which provides a good explanation for the large external shear required by the lens mass model \citep{Hsueh2017}. The monitoring programs of \citet{Koopmans2003a} with the Multi-Element Radio Linked Interferometer Network (MERLIN) and \citet{Rumbaugh2015} with the Jansky Very Large Array (JVLA) showed that CLASS~B0712+472 exhibits indication of intrinsic flux density variability. Keck Adaptive Optics (AO) imaging revealed the presence of an edge-on disc within the lensing galaxy that directly crosses the lensed images A and B \citep{Hsueh2017}. Moreover, the Keck AO imaging also detects the gravitational extended arc related to the emission of the host galaxy of the background AGN.

Images A and B show a flux ratio anomaly, which is an effect usually attributed to the presence of dark matter sub-haloes within the lensing galaxy or along the line of sight \citep{Dalal2002, Keeton2005, Metcalf2005, Despali2018}. However, when the exponential disc component detected in near-infrared imaging is included in the lens mass model, the flux ratios can be fully recovered (\citealt{Hsueh2017}, but see also \citealt{Hsueh2018}). Therefore, in this work we adopt the lens mass model of \citet{Hsueh2017}. 
 
\subsection{CLASS~B1608+656}
CLASS~B1608+656 was also discovered as part of the CLASS survey \citep{Myers2003, Browne2003}. The discovery radio observations of CLASS~B1608+656 revealed four flat-spectrum unresolved images, whilst optical observations with the Hubble Space Telescope ({\it HST}) show arcs created from extended emission of the AGN host galaxy (\citealt{Jackson1998, Koopmans2003b, Suyu2010}, see Fig.~\ref{fig:observations_HST}). The system consists of a source, a post-starburst galaxy located at $z_s$ = 1.394, and two lensing galaxies at $z_l$ = 0.630 \citep{Fassnacht1996}.  CLASS~B1608+656 shows strong flux density variability at both radio and optical wavelengths. Therefore, the system has been identified as an ideal system to measure the Hubble parameter H$_0$ based on gravitationally induced time delay analysis \citep{Fassnacht2002, Koopmans2003b}.  Recently, \citet{Wong2019} estimated H$_0 = 71.0^{+2.0}_{-3.3}$ km s$^{-1}$ Mpc$^{-1}$ from optical monitoring of CLASS~B1608+656. 

All the mass models for this system require a relatively large external shear of $\sim \,10$ per cent, which points to a strong external perturbation from nearby mass.  Indeed, a cluster of galaxies associated with the main lensing galaxies and multiple galaxies along the line of sight have been found \citep{Koopmans2003b, Fassnacht2006, Suyu2010, Greene2013}.  For our analysis, we adopt the model of \citet{Koopmans2003b}, which reproduces well the optical and radio observations, as well as stellar velocity dispersion and the morphology of the Einstein ring detected at optical wavelengths.

\section{Observations}
\label{sec:observations}

In this section, we present new high angular resolution radio and archival optical observations for CLASS~B0712+472 and CLASS~B1608+656, which we use to search for optical--radio offsets.

\subsection{Radio: VLBA observations}
\label{sec:VLBA_observations}

The radio observations of CLASS~B0712+472 and CLASS~B1608+656 were performed with the Very Long Baseline Array (VLBA) at a central observing frequency of 1.65 GHz, with a bandwidth of 256 MHz at 2048 Mbps data rate (Projects ID: BS251, BS257; PI: Spingola). The observation strategy consists of standard phase referencing, which includes scans on the targets of $\sim$5 minutes each interleaved by shorter scans on the phase and bandpass calibrators, for total observation time about 12~h for each observation. The details of the observations are summarised in Table~\ref{tab:VLBA-observations}. The correlation was performed at the VLBA correlator in Socorro and the data were processed with the Astronomical Image Processing Software ({\sc aips}, \citealt{Greisen2003}) package following the standard VLBA calibration procedure for phase-referenced observations, which we summarise\footnote{The details on the calibration procedure can be found in Chaper 9 of {\sc aips} cookbook (\url{http://www.aips.nrao.edu/cook.html}).}.  First, we inspect the visibilities and apply possible flagging in case of bad data, and correct for the ionospheric dispersive delay, which can be significant at low observing frequencies. Then, we correct for voltage offsets in the samplers. After correcting for the instrumental delay and parallactic angle variation, we perform the bandpass calibration using the fringe-finder calibrator and apply \textsl{a priori} amplitude calibration, which uses gain curves and system temperature information. Next, global fringe fitting is performed to correct for the residual fringe delays and rates. Finally, we split out the calibrated visibilities of the target.  

\begin{figure*}
    \centering
        \includegraphics[width = 1.0\textwidth]{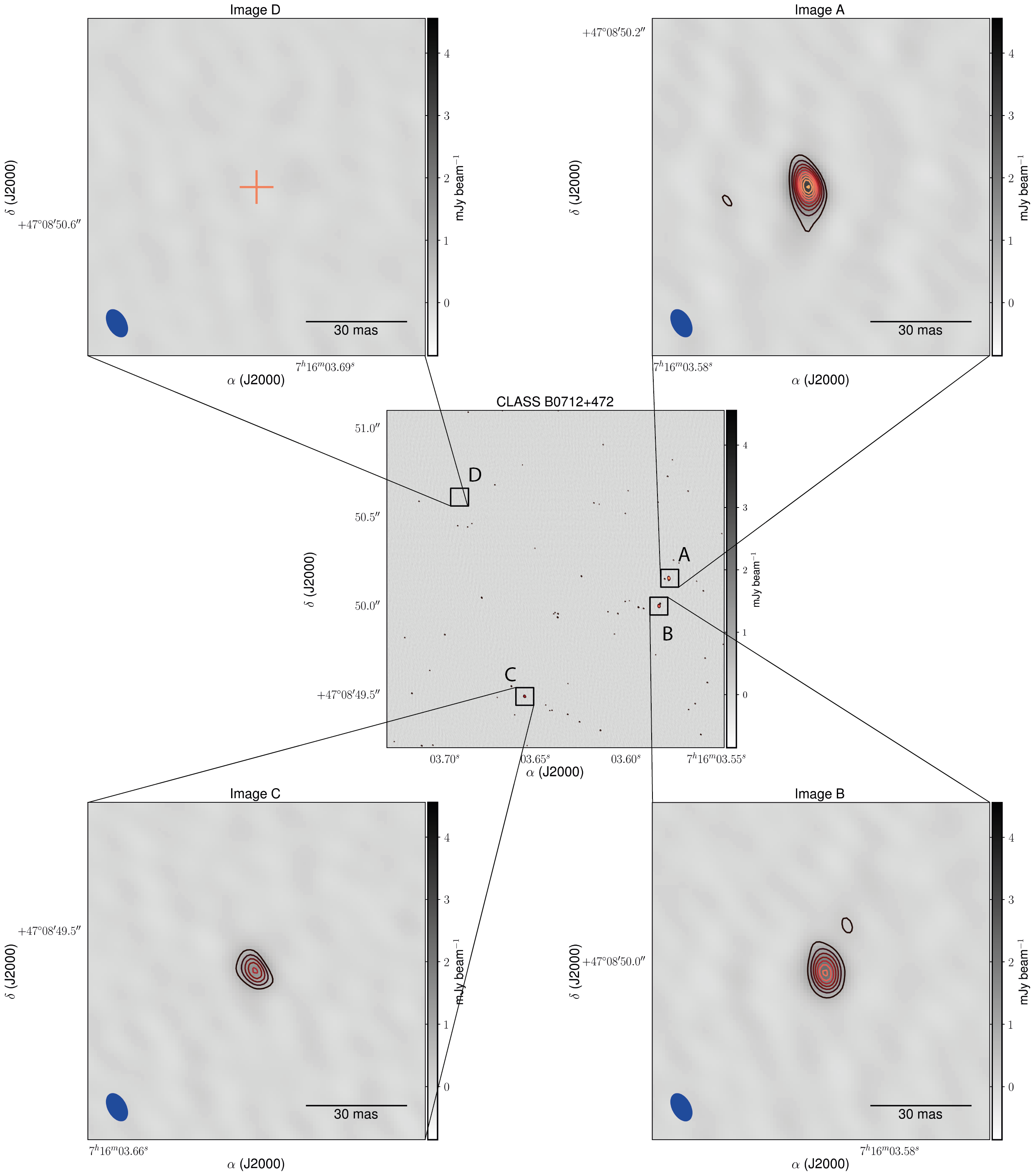}
        \caption{VLBA imaging at 1.7 GHz of the gravitational lens CLASS~B0712+472. The central panel shows the entire system as
observed at 1.7 GHz with VLBI. The grey-scale map is in units of mJy beam$^{-1}$, as indicated by the colour bar on the right of each frame. The image has been obtained using a Briggs weighting scheme with ``robust" = 0. The orange cross indicates the position of image D as estimated using the natural weighted image by \citet{Hsueh2017}. Contours are at (0.1, 0.2, 0.3, 0.4, 0.5, 0.6, 0.7, 0.8, 0.9, 1) $\times$ the peak flux density of each individual image, which is 6 mJy beam$^{-1}$. The off-source rms is about 40 $\mu$Jy beam$^{-1}$. The restoring beam is shown in blue in the bottom left corner and is $9\times5$ mas$^2$ at a position angle of 28.7~deg (east of north). } 
        \label{fig:observations_b0712_VLBI}
\end{figure*}

\begin{figure*}
    \centering
        \includegraphics[width = 1.0\textwidth]{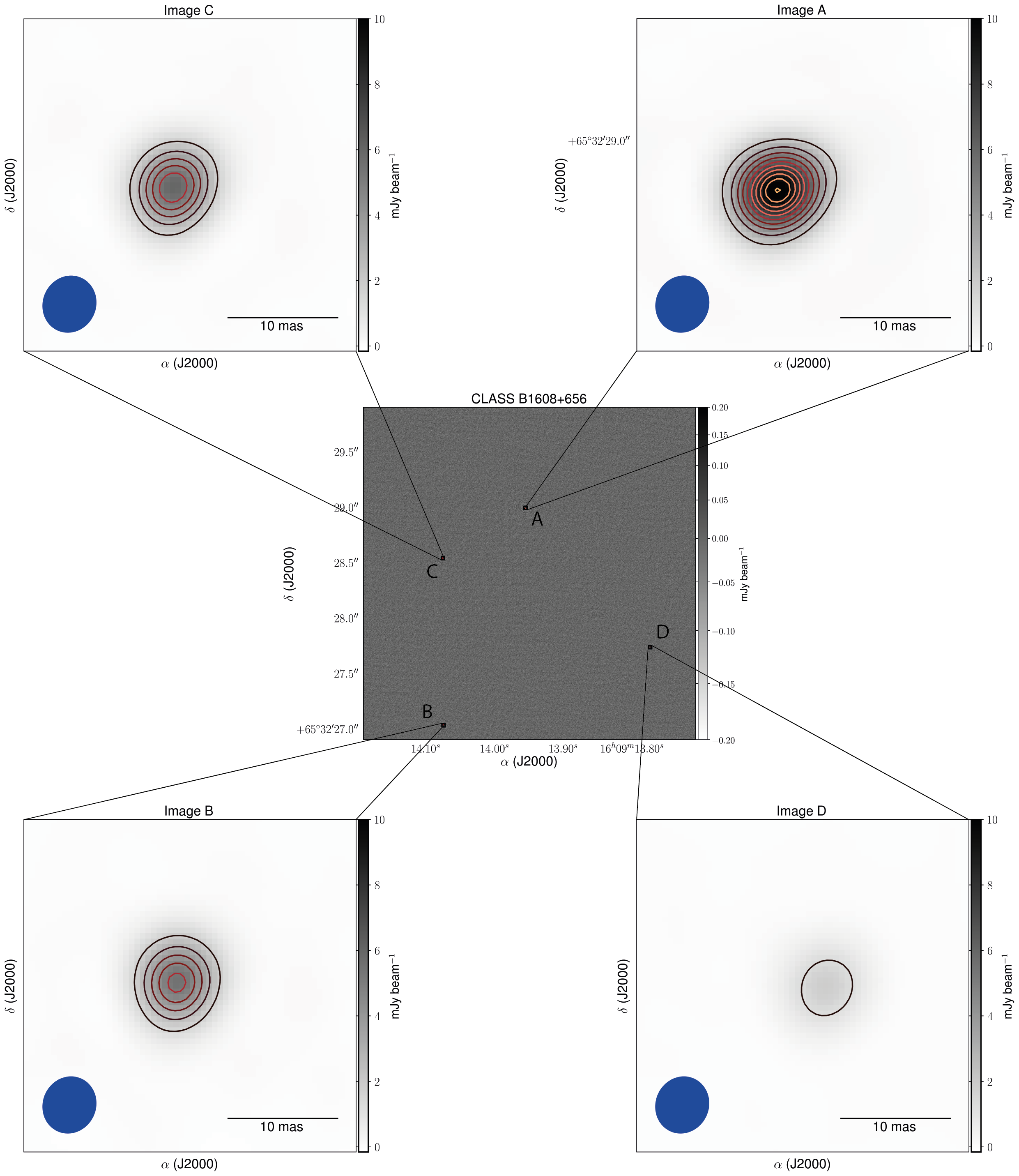}
        \caption{VLBA imaging at 1.7 GHz of the gravitational lens CLASS~B1608+656. The central panel shows the entire system as
observed at 1.7 GHz with VLBI. The grey-scale map is in units of mJy beam$^{-1}$, as indicated by the colour bar on the right of each frame, and we use arcsinh scale to better display the noise structure across the map. The image has been obtained using a Briggs weighting scheme with ``robust" = 0. Contours are at (0.1, 0.2, 0.3, 0.4, 0.5, 0.6, 0.7, 0.8, 0.9, 1) $\times$ the peak flux density of each individual image, which is 16 mJy beam$^{-1}$. The off-source rms is about 25 $\mu$Jy beam$^{-1}$. The restoring beam is shown in blue in the bottom left corner and is $5.3\times4.8$ mas$^2$ at a position angle of 23.9 deg (east of north).} 
        \label{fig:observations_b1608_VLBI}
\end{figure*}

We perform several iterations of phase-only self-calibration starting with a solution interval of 5~min and iteratively decreasing it down to 30~s. The final self-calibrated images of the targets CLASS~B0712+472 and CLASS~B1608+656 are shown in Figs.~\ref{fig:observations_b0712_VLBI} and \ref{fig:observations_b1608_VLBI}, respectively. For the imaging we use the Briggs' weighting scheme with ``robust" parameter equal to 0 \citep{Briggs1995}. 

The self-calibrated image of CLASS~B0712+472 has an off-source rms of 40~$\mu$Jy beam$^{-1}$, a peak flux density of 6 mJy beam$^{-1}$ and the total flux density is $20\pm2$ mJy. The restoring beam is $9\times5$ mas$^2$ at a position angle 28.7~deg (east of north). At this angular resolution only image A is partially resolved, while the other detected lensed images (B and C) are essentially unresolved. 

The off-source rms of the self-calibrated image of CLASS~B1608+656 is 25~$\mu$Jy beam$^{-1}$, the peak flux density is 16~mJy beam$^{-1}$ and the total flux density is $40\pm4$~mJy. The restoring beam is $5.3 \times 4.8$ mas$^2$ at a position angle $−$23.9~deg (east of north). This is the highest angular resolution VLBI image to date for this system. At this angular resolution all of the lensed images of this system are mainly unresolved, except for image A, which shows an indication for a possible extended structure (see Fig.~\ref{fig:observations_b1608_VLBI}).

The position of the lensed images of both systems are measured from a Gaussian fit to the observed emission in the image-plane by using the task {\tt jmfit} within {\sc AIPS}, and they are listed in Tables~\ref{tab:b0712_positions} and \ref{tab:b1608_positions} for CLASS~B0712+472 and CLASS~B1608+656, respectively. The uncertainty on the position is estimated using standard formulae from \citealt{Condon1997}, by taking into account the rms of the self-calibrated image and the peak surface brightness of each lensed image.

\begin{table*}
\caption{Summary of the VLBA observations. }
\label{tab:VLBA-observations}
\begin{tabular}{llllllllll}
\hline
Target & ID & Obs. date & Bandwidth & t$_{\rm exp}$ (h) & IF & Correlations \\
\hline
CLASS~B0712+472 & BS251 &  2016 Feb 13 & 256 MHz & 7.6 & 8 & RR, LL\\
CLASS~B1608+656 & BS257 &  2017 Jun 3 & 256 MHz & 7.8 & 8 & RR, LL\\

\hline
\end{tabular}
\end{table*}

\begin{figure*}
    \centering
   \includegraphics[width = 0.98\textwidth]{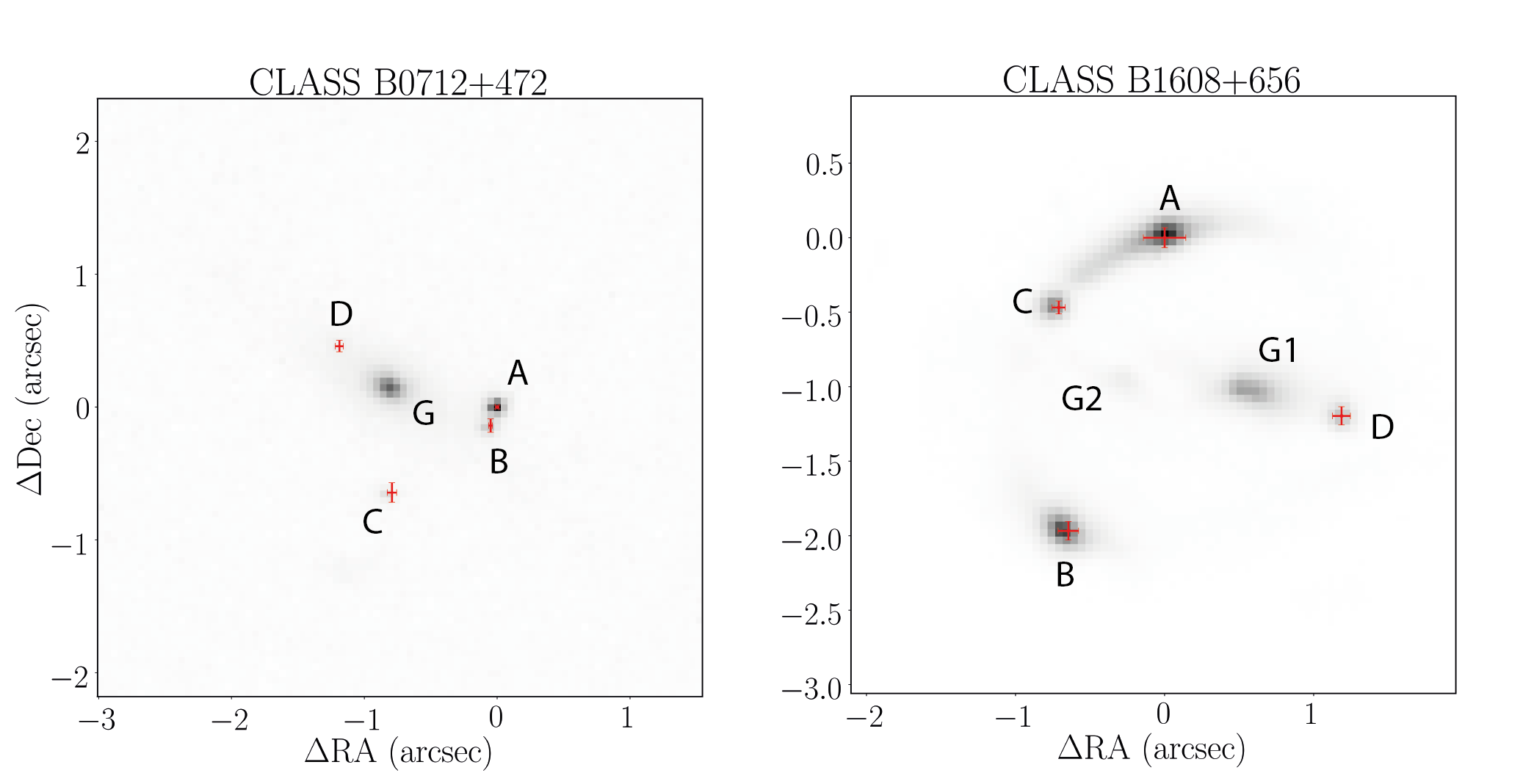}
    \caption{\textsl{Left:} Optical imaging of CLASS~B0712+472 taken with the {\it HST}-WFPC2/F555W filter. The red dots and errorbars indicate the position of the optical lensed images and their uncertainties as estimated using the method explained in Sec.~\ref{sec:observations}. The letters A, B, C and D label the lensed images, while G indicates the lensing galaxy, which is an edge-on spiral galaxy for this system. \textsl{Right:} Optical imaging of CLASS~B1608+656 using the {\it HST}-WFPC2/F606W filter, same legend. The labels G1 and G2 indicate the two lensing galaxies. North is up, east is left.} 
    \label{fig:observations_HST}
\end{figure*}

\subsection{Optical: \textit{HST} observations}
\label{sec:hst_observations}

 We use archival {\it HST} data for both systems. CLASS~B0712+472 was observed with the {\it HST} using the Wide Field Planetary Camera 2 (WFPC2) at F555W filter (Program ID: 9138), while CLASS~B1608+656 was observed at F606W filter (Program ID: 10158). We processed the data using standard procedures in {\sc multidrizzle} and show the final images in Fig.~\ref{fig:observations_HST} (see also \citealt{Koopmans2003b, Hsueh2017}). Both of the lensed sources are in fold configuration, showing compact (CLASS~B0712+472) and extended (CLASS~B1608+656) emission. In particular, the lensed images in CLASS~B1608+656 are stretched in the tangential direction and form extended gravitational arcs.  We proceed as follows to account for the extended emission of the lensed images observed at optical wavelengths.
 
 We extract the position of the lensed images using {\tt Photutils} package\footnote{We refer to the {\tt Photutils} documentation for additional information \url{https://photutils.readthedocs.io/en/stable/segmentation.html}.} \citep{Bradley2019}, which includes general-use functions to detect sources (both point-like and extended) in an image using a process called \textsl{image segmentation}, which consists of assigning a label to every pixel in an image such that pixels with the same label are part of the same source. The segmentation procedure implemented in {\tt Photutils} is called the \textsl{threshold method}, where detected sources must have a minimum number of connected pixels that are each greater than a specified threshold value in an image. The threshold level is usually defined at some multiple of the background noise (sigma) above the background. We have used $3\, \sigma$ as the threshold level. The image can also be filtered before thresholding to smooth the noise and maximise the detectability of objects with a shape similar to the filter kernel. Moreover, the lensed images observed at optical wavelengths can be blended with the emission of the lensing galaxy. Therefore, we de-blend the sources using the {\tt Photutils} methods that combine multi-thresholding and watershed segmentation. 

The results for CLASS~B0712+472 and CLASS~B1608+656 are summarised (as ``optical extended") in Tables ~\ref{tab:b0712_positions} and \ref{tab:b1608_positions}, respectively, and are shown in Figure~\ref{fig:observations_HST}. This approach provides the position of the lensed images with sub-pixel accuracy when the images are well represented by a Gaussian profile. However, we reiterate that the brightest lensed images are subjected to the strongest magnification in the tangential direction, which results in arc-like stretched shapes (see Fig.~\ref{fig:observations_HST}). For this reason, the uncertainty of the position of the brightest lensed images of B1608+656 (A and B) is larger than the other two images.
To better estimate the position of the peak of the emission, we explore a multi-dimensional image processing method using {\tt scipy.ndimage} which provides the position of the brightest pixel \footnote{We refer to \url{https://docs.scipy.org/doc/scipy/reference/ndimage.html} for further details on {\tt scipy.ndimage}.} and {\tt scipy.ndimage.filters.maximum\_filter()}\footnote{We refer to \url{https://docs.scipy.org/doc/scipy/reference/generated/scipy.ndimage.maximum_filter.html} for further information on {\tt scipy.ndimage.filters.maximum\_filter()} method.}. 
Using this method, the positional uncertainties are at the pixel scale-level also for a highly asymmetric profile of the lensed images, and the results are summarised as ``optical point" in Tables~\ref{tab:b0712_positions} and \ref{tab:b1608_positions} for CLASS~B0712+472 and CLASS~B1608+656, respectively.

\begin{table*}
\caption{Summary of the {\it HST} observations.}
\label{tab:HST-observations}
\begin{tabular}{lllllll}
\hline
Target & Proposal ID & Obs. date & Detector & Filter & t$_{\rm exp}$ (h)\\
\hline
CLASS~B0712+472 		& 9138  & 2001 Oct 27 & WFPC2 & F555W & 2.2\\
CLASS~B1608+656 		& 10158 & 2004 Aug 29 & WFPC2 & F606W & 0.72\\
\hline
\end{tabular}
\end{table*}

\section{Source inversion}
\label{sec:source_inversion}
In this section, we focus on the method applied for the multi-wavelength source reconstruction and  analysis of its positional uncertainty.

\subsection{Lens mass model}
\label{sec:lensmodelling}

In order to reconstruct the radio and optical emissions of both targets we adopt a parametric approach, and use the publicly available software {\sc gravlens} \citep{Keeton2001}. We adopt a backward-ray tracing method: namely, we fix parameters of the lens mass model to values from \citet{Hsueh2017} and \citet{Koopmans2003b} for CLASS~B0712+472 and CLASS~B1608+656, respectively. The mass models used for our analysis are summarised in Table~\ref{Tab:lensmodels}. We highlight that the radio and optical images are aligned to a common reference frame and, therefore, also the reconstructed radio and optical sources are aligned on the same coordinate system.

We use Markov-Chain Monte Carlo (MCMC) sampler implemented in {\sc gravlens} (Table~\ref{Tab:lensmodels}) to estimate the uncertainties (68 per cent confidence level) on the lens mass model parameters of CLASS~B1608+656. The lens models for CLASS~B0712+472 and CLASS~B1608+656 reproduce the positions of the lensed images within the uncertainties for both the radio and optical observations (see Table~\ref{tab:b0712_positions} and \ref{tab:b1608_positions}). This indicates that the model represents the data well. 
We only use the position of the lensed images (estimated using the method described in Sec.~\ref{sec:observations}) to infer the position of the source. We do not use flux ratios, as they could be affected by multiple phenomena, such as  microlensing at optical wavelengths, intrinsic variability of the radio source (observed in both systems), substructure along the line-of-sight, and extrinsic variability due to the presence of the interstellar medium \citep{Koopmans2003a, Rumbaugh2015, Despali2018, Vernardos2019}. 

In the case of CLASS~B0712+472, image D is not detected using the weighting Briggs' weighting scheme with ``robust" = 0 (Fig.~\ref{fig:observations_b0712_VLBI}). However, the choice of this weighting of the visibilities results in a smaller restoring beam with respect to a natural weighting and, therefore, gives better positional uncertainty on the other lensed images. For this reason, we use the position of image D estimated using natural weights in \citet{Hsueh2017} and our estimate for the other lensed images.

For the purpose of relative astrometry, we assign coordinates (0,0) arcsec to the VLBA lensed image A. Using a lensed image as reference does not prevent us from detecting any possible offset between the radio and optical emissions, because any optical--radio positional mismatch would be visible in the other lensed images. As a further check, we also performed the backward-ray tracing analysis by using the VLBA lensed image B as reference, as the choice of the reference image is expected to not affect the recovered relative optical--radio offset in the source plane. Indeed, we find that the VLBI--optical offsets using image B (VLBA) as reference are completely consistent with what found by using VLBA-image A as reference image.

%------------------- Table position of B0712+472 components -------------------
 \begin{table*}
  \centering
  \caption{Position of the lensed images of the system CLASS~B0712+472. The observed relative positions ($\Delta x$, $\Delta y$) are determined for the radio data by performing elliptical Gaussian fits using {\sc jmfit} within {\sc aips}, and are measured with respect to the VLBA component A (phased referenced absolute position 07$^h$16$^m$03.576$^s$, +47$^{\circ}$08'50.154"). The optical positions are determined as explained in Section \ref{sec:hst_observations}. The predicted positions from the lens model are also given relatively to the VLBA component A. }\label{tab:b0712_positions} 
	\begin{tabular}{lllllll}
    	\hline
		\noalign{\vskip 0.15cm}    
			\multirow{2}{*}{Band} & \multirow{2}{*}{ID} & \multicolumn{2}{c}{Observed} & \multicolumn{2}{c}{Model} \\
		\noalign{\vskip 0.05cm} 
	 	& & $\Delta x$ (arcsec) & $\Delta y$ (arcsec)  & $\Delta x^{\rm mod}$ (arcsec) & $\Delta y^{\rm mod}$ (arcsec) \\
        \hline
		\noalign{\vskip 0.15cm}    
		\multirow{4}{*}{Radio} & A &  $\equiv$0.0000$\pm$0.0003 &  $\equiv$0.0000$\pm$0.0001 & $+$0.0001 &  $-$0.0002 \\
        & B & $-$0.0562$\pm$0.0003 & $-$0.1583$\pm$0.0002 &  $-$0.0565  &  $-$0.1582  \\
        & C & $-$0.8123$\pm$0.0003 & $-$0.6664$\pm$0.0002 & $-$0.8122 &  $-$0.6666 \\
        & D & $-$1.1741$\pm$0.015  & $+$0.4593$\pm$0.003  &  $-$1.1739 &  $+$0.4591  \\ [6pt]	
	   
	   \multirow{4}{*}{Optical (extended)}	
	   &    A	& 0.000$\pm$0.014	    & 0.000$\pm$0.014	        &  $+$0.001 &  $+$0.011	 \\
	   &	B	& $-$0.0495$\pm$0.0506  & $-$0.1395$\pm$0.0142		&  $-$0.0493   & $-$0.1382	 \\
	   &	C	&  $-$0.7920$\pm$0.0737 & $-$0.6435$\pm$0.0349	    &  $-$0.7912	& $-$0.6413  \\
	   &    D	&  $-$1.188$\pm$0.043  & $+$0.4590$\pm$0.0302 	    &  $-$1.145	& $+$0.4583		\\ [6pt]	
     	\noalign{\vskip 0.1cm}  
 
     \multirow{4}{*}{Optical (point)}	
     &  A	& 0.000$\pm$0.025			& 0.000$\pm$0.025	 		&  $-$0.003	& $-$0.013	\\
	 & 	B	& $-$0.050$\pm$0.025	    & $-$0.149$\pm$0.025		&  $-$0.050  	& $-$0.149	 \\
	 &	C	& $-$0.792$\pm$0.025      	& $-$0.644$\pm$0.025	   	&  $-$0.819	& $-$0.645	 \\
	 &  D	& $-$1.188$\pm0.025$		& $+$0.446$\pm$0.025	   	&  $-$1.203 	& $+$0.460	\\ [6pt]	
     	\noalign{\vskip 0.1cm} 
     	
     	\hline
	\end{tabular} 
\end{table*} 
%------------------- End Table B0712 positions ------------------------------

%------------------- Table position of B1608+656 components -------------------
 \begin{table*}
  \centering
  \caption{Position of the lensed images of the system CLASS~B0712+472. The observed relative positions ($\Delta x$, $\Delta y$) are determined for the radio data by performing elliptical Gaussian fits using {\sc jmfit} within {\sc aips}, and are measured with respect to the VLBA component A (phased referenced absolute position 16$^h$09$^m$13.956$^s$, +65$^{\circ}$32'28.995"). The optical positions are determined as explained in Section \ref{sec:hst_observations}. The predicted positions from the lens model are also given relatively to the VLBA component A.}\label{tab:b1608_positions} 
	\begin{tabular}{lllllll}
    	\hline
		\noalign{\vskip 0.15cm}    
			\multirow{2}{*}{Band} & \multirow{2}{*}{ID} & \multicolumn{2}{c}{Observed} & \multicolumn{2}{c}{Model} \\
		\noalign{\vskip 0.05cm} 
	 	& & $\Delta x$ (arcsec) & $\Delta y$ (arcsec) & $\Delta x^{\rm mod}$ (arcsec)  &  $\Delta y^{\rm mod}$ (arcsec)  \\
        \hline
	    \noalign{\vskip 0.15cm}    
		\multirow{4}{*}{Radio}	& A	& $\equiv$0.0000$\pm$0.00003 &  $\equiv$0.0000$\pm$0.00004	& $+$0.00003 & $+$0.00000 \\
	    &	B	& $-$0.73781 $\pm$0.00003 & $-$1.96098$\pm$0.00004 &  $-$0.73780	& $-$1.96102 \\
	    &	C	& $-$0.75774 $\pm$0.00004 & $-$0.45396$\pm$0.00005	&  $-$0.75777 & $-$0.45393	 \\
	    &   D	& $+$1.13989$\pm$0.00005  & $-$1.25630$\pm$0.00006  &  $+$1.13990 &  $-$1.25630 \\ [6pt]	

	\multirow{4}{*}{Optical (extended)}	&   A	&  $-$0.003$\pm$0.143		&  $+$0.003$\pm$0.067		& $+$0.095	& $+$0.036	 \\
     & B & $-$0.649$\pm$0.069   &  $-$1.964$\pm$0.062		& $-$0.688 		& $-$1.912	\\
     & C & $-$0.714$\pm$0.045   & $-$0.465$\pm$0.043	   	&  $-$0.746		& $-$0.448	\\
	 & D &  $+$1.184$\pm$0.060	 & $-$1.193$\pm$0.061	   	& $+$1.242	   & $-$1.256	\\ [6pt]	
     	\noalign{\vskip 0.1cm}  
 
     \multirow{4}{*}{Optical (point)}	& A & $-$0.003$\pm$0.025   &  $+$0.003$\pm$0.025 		& $-$0.010		& $+$0.004 \\
    & B	&  $-$0.654$\pm$0.025    &  $-$1.986$\pm$0.025    &  $-$0.671  &  $-$1.978	\\
    & C	& $-$0.753$\pm$0.025    & $-$0.486$\pm$0.025   &  $-$0.753  &  $-$0.505	\\
    & D	&  $+$1.146$\pm$0.025    & $-$1.186$\pm$0.025   &  $+$1.160  & $-$1.190	\\ [6pt]	
     	\noalign{\vskip 0.1cm}  
     	\hline
	\end{tabular} 
\end{table*} 
%------------------- End Table B1608 positions ------------------------------

 %--------------Table Lens mass model parameters
   \begin{table*}
      \centering
      \caption{Parameters of the lens model for the mass density distribution of the main deflectors in CLASS~B0712+472 and CLASS~B1608+656. CLASS~B0712+472 has one lensing galaxy, which is parametrised by two mass components (an elliptical power-law plus an exponential disk), while in CLASS~B1608+656 there are two lensing galaxies acting as lenses, which are parametrised as two elliptical power-laws. The mass models assumed here are those from \citet{Hsueh2017} and \citet{Koopmans2003b}, where the mass strength $b$ and the scale radius $R_S$ are in arcsec; the position angles are in degrees (east of north). The density slope is given by $\alpha$, where $\alpha = 1$ corresponds to an isothermal profile \citep{Keeton2001}. The lensing galaxy position ($\Delta x$, $\Delta y$) is given for both systems with respect to VLBA-image A.}
         \label{Tab:lensmodels}
  \begin{tabular}{lcccc}
  	\hline
  Parameter & \multicolumn{2}{c}{CLASS~B0712+472}  &  \multicolumn{2}{c}{CLASS~B1608+656} \\
  & PL & expdisk & G1 (PL) & G2 (PL)  \\
    \hline
        $b$ & 0.609$\pm$0.007 & &  0.531$\pm0.008$ & 0.288$\pm0.002$ \\
         $\Delta x$ & $+$0.785$\pm$0.005 & $+$0.896$\pm$0.005 & $+$0.425$\pm0.009$ &$-$0.291$\pm$0.008\\
        $\Delta y$ & $+$0.142$\pm$0.004  & $+$0.200$\pm$0.004 & $-$1.071$\pm0.007$ & $-$0.929$\pm0.002$ \\
        $e$ & 0.16$\pm$0.01  &  0.77$\pm$0.01 & 0.4$\pm0.1$ & 0.7$\pm$0.1\\
        $\vartheta$ & $+$71$\pm$3 & $+$59.7$\pm$0.3 & 77$\pm1$ & 69.2$\pm$0.1\\
	    $\Gamma$ &  0.096$\pm$0.005  &  & 0.08$\pm$0.01 &   \\
        $\Gamma_{\vartheta}$ & 34.4$\pm$1.6 & &  7$\pm2$ &  \\
        $\alpha$ & $\equiv$1&  & $\equiv$1 & $\equiv$1 \\ 
        $\kappa_0$ & & 0.29$\pm$0.04 & & \\
        $R_S$ & & 0.389$\pm$0.02 & & \\
    \hline
        
    \end{tabular}
	\end{table*}

\subsection{Uncertainty on the recovered source position}
\label{sec:uncertainty_source}

We use Monte Carlo simulations to estimate the uncertainty of the reconstructed source position at radio and optical wavelengths following the same approach explained in \citet{Spingola2019c}. We briefly outline the approach below.

We start by generating mock lensed images based on Gaussian sampling. The Gaussian distributions that we sample from is centered at the observed position of the lensed images and the standard deviation corresponding to its uncertainty (see Table \ref{tab:b0712_positions} and Table~\ref{tab:b1608_positions}).We generate mock lens model parameters within the uncertainties given by the MCMC sampler implemented in {\sc gravlens} (Table \ref{Tab:lensmodels}) to also incorporate effect of the uncertainties in the lens model on the source location. Next, we backward ray-trace the mock lensed images produced using the mock lens models to obtain many realizations for each source component (radio and optical). Finally, the uncertainty of each source component is estimated as the standard deviation of the backward ray-traced source positions. 

\section{Results}
\label{sec:results}
\begin{figure*}
\centering
	\includegraphics[width = 0.95\textwidth]{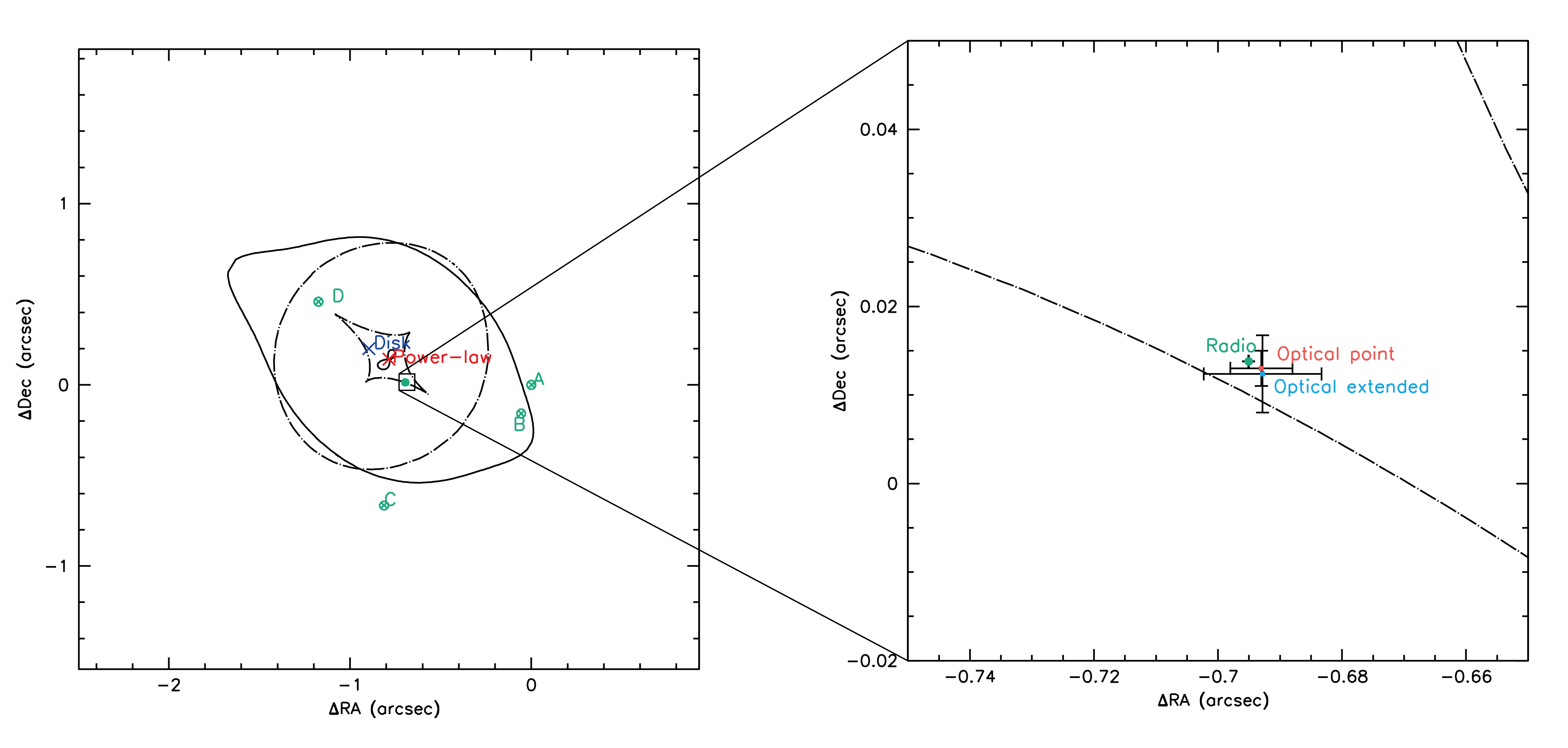}
    \caption{\textsl{Left:} Lens mass model for CLASS~B0712+472. The observed radio positions are the open circles and the predicted positions are represented by the crosses, with all positions given relative to component A. The filled circle represents the source. The lens critical curve is shown by the solid line, while the source plane caustics are indicated by the dashed line. The red cross indicates the position of the centroid of the power-law mass density distribution component, while the blue cross indicates the centroid of the exponential disk component. \textsl{Right:} Source plane reconstruction of the optical extended (blue), optical point(red) and radio (green) emission within CLASS~B0712+472.}
    \label{Fig:model_b0712}
    \end{figure*}

In this section, we summarise the results of the backward ray-tracing analysis described in Sec.~\ref{sec:source_inversion} for our two targets. 
Figs.~\ref{Fig:model_b0712} and ~\ref{Fig:model_b1608} show the lens and source planes of CLASS~B0712+472 and CLASS~B1608+656 with critical and caustic curves. The position of the optical and radio components in the source plane, shown in Figs. ~\ref{Fig:model_b0712} and ~\ref{Fig:model_b1608}, are also listed in Table~\ref{tab:source}.

\subsection{CLASS~B0712+472}
 
The source plane reconstruction of CLASS~B0712+472 finds that the optical and radio emissions are co-spatial within their positional uncertainties, with relative offset $2\pm9$~mas. In physical units, this corresponds to a relative positional offset of $17\pm77$~pc\footnote{At a redshift $z=1.34$, 1 mas corresponds to 8.536~pc.}. When using the uncertainties on the optical position obtained with the multi--dimensional image processing method (listed in Table~\ref{tab:b0712_positions}), we find that the offset between the radio and optical sources is of $2\pm5$ mas, which corresponds to $17 \pm 42$~pc (see Table~\ref{sec:source_inversion}). Therefore, we can conclude that the optical and radio emissions in CLASS~B0712+472 spatially coincide within the central $\sim40\,$pc. 
This measurement demonstrates that this approach can provide orders of magnitude of improvements in spatial resolution with respect to observations of un-lensed sources, and constrain the relative position of multi-wavelength emissions at pc-level in sources at redshift even greater than one. 

\subsection{CLASS~B1608+656}
 
\begin{figure*}
\centering
    \includegraphics[width = 0.95\textwidth]{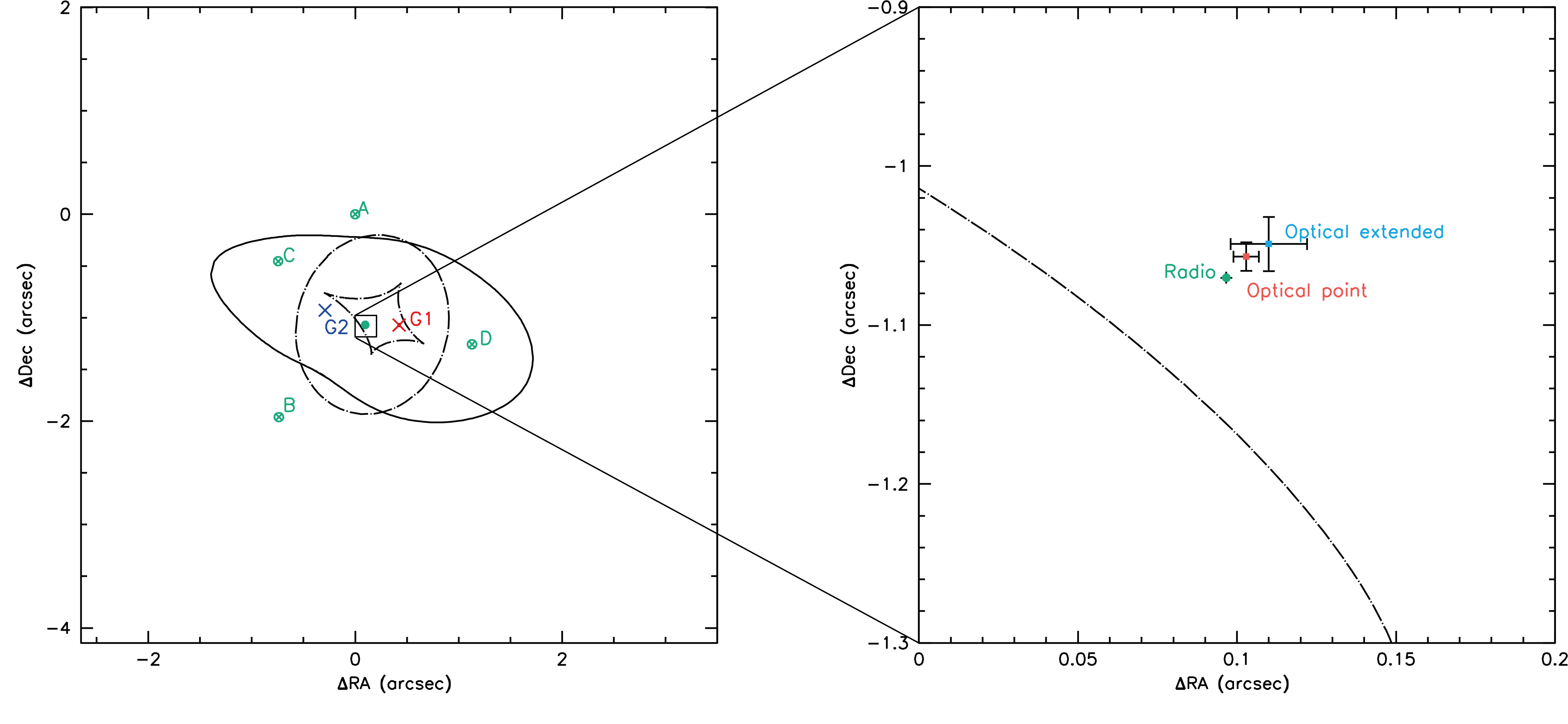}
    \caption{\textsl{Left:} Lens mass model for CLASS~B1608+656. 
    The observed radio positions are the open circles and the predicted positions are represented by the crosses, with all positions given relative to component A. The filled circle represents the source. 
    The lens critical curve is shown by the grey dots, while the source plane caustics are indicated by the black dots. 
    The red cross indicates the position of the centroid of the power-law mass density distribution component associated with the main lensing galaxy G1, while the blue cross indicates the centroid of the power-law mass density distribution component associated with satellite galaxy G2. 
    \textsl{Right:} Source plane reconstruction of the optical extended (blue), optical point (red) and radio (green) emission within CLASS~B1608+656.}
    \label{Fig:model_b1608}
    \end{figure*}

The observations of CLASS~B1608+656 show significant differences in morphology of the lensed images at radio and optical wavelengths, with the radio emission being compact whilst the optical demonstrate extended arcs (see Figures~\ref{fig:observations_b1608_VLBI} and~\ref{Fig:model_b1608}). In this system our source reconstruction finds evidence for an offset between the compact radio emission and the center of extended optical emission. The estimated offset is $25\pm16$~mas (also shown in Fig.~\ref{Fig:model_b1608}). At $z=1.394$, this offset corresponds to $214\pm137$~pc\footnote{At a redshift $z=1.394$, 1 mas corresponds to 8.565~pc}. 
However, the point optical emission traced by the brightest pixel and the VLBI component provide an offset of $14\pm 8\,$mas, which at redshift of the source of 1.394, corresponds to $120\pm 69\,$pc.

As shown also in Fig.~\ref{Fig:model_b1608}, the optical emission given by the brightest pixel seems to be closer to the radio emission with an offset of $\sim 40$~pc (though consistent with the other estimate within the uncertainties, as expected), while the centroid of the extended optical emission is further away at $\sim200 \,$pc distance to the VLBI radio core.
Interestingly, CLASS~B1608+656 has been spectroscopically identified as a post-starburst galaxy \citep{Fassnacht1998}. 
Therefore, the measured offset between the compact and extended emission, combined with the evolutionary phase of the host galaxy makes CLASS~B1608+656 an excellent candidate for being an offset--AGN. This aspect will be further discussed in Sec.~\ref{sec:B1608+656OffsetAGN}.
 
Our approach proves that gravitational lensing can open a new path towards the identification of offset--AGN candidates at redshifts greater than one, and, as such, can give us an unprecedented way to shed new light into galaxy evolution. 
 
  %--------------------- Table of source position ---------------------

\begin{table*}	
	\centering
	\caption{Positions of the radio and optical source components relative to lensed image A (VLBA), which is at (0,0)~arcsec (see also Tables~\ref{tab:b0712_positions} and \ref{tab:b1608_positions}). The positional offset between the optical and radio sources is estimated by using the optical ``extended" and ``point" source, as explained in Sec.~\ref{sec:hst_observations}, and is given in arcseconds and parsecs. The uncertainties listed are statistical uncertainties, which take into account both the observational and lens mass model uncertainties. Redshifts are from \citet{Fassnacht1998} and \citet{Fassnacht1996} for CLASS~B0712+472 and CLASS~B1608+656, respectively. In the case of CLASS~B0712+472, the emission at optical and radio wavelengths is co-spatial within the uncertainties. In the source CLASS~B1608+656 there is an offset between the radio and optical components.}  
	\label{tab:source}
	\resizebox{0.999\textwidth}{!}{%
	\begin{tabular}{lllllll} 
	\hline
	Target  & Source & $\Delta x$ (arcsec) & $\Delta y$ (arcsec)  & optical--radio offset (arcsec) & optical--radio offset (pc)  & z$_s$\\
   \hline
    \multirow{3}{*}{CLASS~B0712+472} & Radio & $-$0.69507$\pm$0.00005 & $+$0.01380$\pm$0.00002 &   \multirow{2}{*}{0.002$\pm$0.009}& \multirow{2}{*}{17$\pm$77} &  \multirow{3}{*}{1.34}  \\
      & Optical (extended) & $-$0.693$\pm$0.010 & $+$0.013$\pm$0.004\\
      & Optical (point)   & $-$0.693$\pm$0.005 & $+$0.013$\pm$0.002 & 0.002$\pm$0.005 & 17$\pm$42 \\ 

 	\hline
    \multirow{3}{*}{CLASS~B1608+656} & Radio &  0.096638$\pm$0.000005 & $-$1.070248$\pm$0.000009 & \multirow{2}{*}{0.025$\pm$0.016} & \multirow{2}{*}{214$\pm$137}& \multirow{3}{*}{1.394}\\
     & Optical (extended) &  0.110$\pm$0.012 & $-$1.049$\pm$0.017 \\
     & Optical (point)   & 0.102$\pm$0.004  &  $-$1.057$\pm$0.009 &  0.014$\pm$0.008 & 120$\pm$69\\ 
     \hline
\end{tabular}
}
\end{table*}

%--------------------- Figure B0712+472 model ---------------------
    
%--------------------- Figure B1608+656 model ---------------------
    
\section{Discussion}
\label{sec:discussion}
    
The central regions of active galaxies provide a unique laboratory to study galaxy evolution and, in particular, their formation history \citep{Kormendy1992,Conselice2007}. However, the study of multi-wavelength emission of AGN at scales smaller than $200\,$pc is severely limited by the angular resolution and astrometric precision of current observatories. Our results demonstrate the capability  of combining the highest angular resolution observations available with gravitational lensing to spatially resolve the multi-wavelength radiation from hundreds of parsecs down to even $0.08\,$pc (in the case of CLASS~B1608+656).

In this Section, we discuss the precision of our method applied to CLASS~B0712+472 and CLASS~B1608+656 and the path towards further improvements (Sec.~\ref{Sec:astrometric_precision}), the nature of CLASS~B1608+656 as an offset--AGN candidate (Sec.~\ref{sec:B1608+656OffsetAGN}), and its impact on the entire offset--AGN population (Sec.~\ref{sec:PopulationOffsetAGN}). 
Finally, we discuss our method and results as a probe of the cosmological evolution of galaxies, as well as the application of this methodology to the upcoming multi-wavelength surveys (Sec.~\ref{sec:GalaxyEvolution_LargeSample}).

\subsection{Limitations of the precision of the method} 
\label{Sec:astrometric_precision}

\subsubsection{Radio VLBI observations} 
  
We combine in our analysis strong gravitational lensing with milliarcsecond angular resolution VLBI observations. As the morphology of lensed images is mostly unresolved, this combination allows us to put an upper limit on the size of the radio emission, which is $\leq$0.08~pc in CLASS~B1608+656 and $\leq$0.4~pc in CLASS~B0712+472 (Table ~\ref{tab:source}).

The gain in astrometric precision on the radio emission from gravitational lensing is of at least an order of magnitude, from the $\sim0.3-2.5$~mas precision on the lensed images position to the $5-50$~$\mu$as precision in reconstructing the position of the radio source (Tables \ref{tab:b0712_positions}, \ref{tab:b1608_positions} and \ref{tab:source}). For comparison, diffraction-limited resolution of the Event Horizon Telescope (EHT) is $\sim25$~$\mu$as \citep{EHT}.
To achieve such improvement in astrometric precision at the same observing frequency without the lensing boosting, it would be necessary to increase the maximum baseline of the radio interferometer to $\sim 12000$ km (the maximum baseline of the VLBA is $\sim 8000$ km), assuming that the source is unresolved also at those baselines. Otherwise, it would also require a much longer observing time to detect the source at the same signal-to-noise ratio of the ``lower" angular resolution observations.

\subsubsection{Optical observations} 

The uncertainty of the optical source position of the CLASS~B0712+472 system 
is about $10$~mas. Such large difference with respect to radio precision is expected given the two orders of magnitude difference in angular resolution of the VLBI observations compared to that of {\it HST}.
We find that the optical positional uncertainties obtained using a multi--dimensional image processing method (indicated as ``optical point" in Tables~\ref{tab:b0712_positions} and ~\ref{tab:source}) improve by a factor of two with respect to those obtained by taking into account the extended emission of the lensed images.

The relative positions of point-like sources of HST images can be measured with sub-mas accuracy \citep{2011PASP..123..622B}. 
As demonstrated by \citet{2018ApJ...855..136R}, spatial scanning with HST's WFC3 can provide relative astrometry with $30-40$ $\mu$as precision. Therefore, applying more elaborated methods to optical observations of lensed sources in the caustic configuration could provide further improvements of localization of optical emissions in relation to radio sources.

\subsubsection{Lens modelling} 

As the caustics in a gravitational potential of lensing galaxies act as non-linear amplifiers, the created multiple lensed images provide strong constraints on the model of the lens  \citep[e.g.][]{Barnacka2017,Barnacka2018}. A precise lens model is critical for reconstructing the position of the source with high accuracy.  
For the systems CLASS~B0712+472 and CLASS~B1608+656, the uncertainties of the lens mass model parameters are about $10$~per cent (see Table~\ref{Tab:lensmodels}). This high precision of the model parameters is mainly due to the strong constraints of the mass model, which are set by the high angular resolution and sensitivity of the VLBI observations. 

The lens mass model can be substantially improved if the lensed sources show extended emission, as in the case when, for example relativistic jets form extended arcs or multiple-lensed images. 
Such complex sources provide observations that can tightly constrain the tangential direction of the gravitational potential. If the multiple lensed images also show extended emissions towards the center of the lens, then the radial structure of the gravitational potential can also be highly constrained. In this case, when there are strong constraints for both radial and tangential direction, it is possible to infer the lens mass model parameters at sub-per cent level, as for example for the extended gravitational arcs in MG~J0751+2716 \citep[e.g.][]{Spingola2018}. Also, in  HS~0810+2554, a radio-loud gravitationally lensed system, observed with VLBI it was possible to infer the lens mass model parameters at a sub-per cent level, as the eight lensed images sampled well the tangential and radial direction of the gravitational potential \citep{Hartley2019}. 

\subsection{CLASS~ B1608+656 as an offset-AGN candidate}
\label{sec:B1608+656OffsetAGN}

We use the properties of gravitational lensing to investigate the spatial location of the optical and radio emissions in distant galaxies with high positional precision. 

Our results for CLASS~B1608+656 reveal an offset of $214\pm137$~pc between the optical and radio emissions in Fig.~\ref{Fig:model_b1608} and Table~\ref{tab:source}. As the lensed images observed with {\it HST} show significant extent, the optical emission is likely indicating the stellar light from an AGN host galaxy. On the other hand, the VLBI observations show high level of compactness in the lensed images (as they are unresolved at 5~mas angular resolution), which can be associated with a central AGN component. 

An optical--radio offset does not always imply the presence of an offset--AGN. For example, optical--radio offsets could be associated with compact symmetric objects (CSO), as for the gravitationally lensed sources HS~0810+2554 and JVAS~B1938+666, where two flat-spectrum radio components are detected \citep{Hartley2019, Spingola2019sharp}. However, there is no evidence for multiple flat-spectrum radio components in the background source of CLASS~B1608+656 down to the high angular resolution (5~mas) provided by our VLBI observations (see also Fig.~\ref{fig:observations_b1608_VLBI}). Optical--radio offsets can be also due to the presence of a pc-scale jet that is generally unresolved at optical wavelengths \citep{Kovalev2017, Petrov2017, Petrov2019}. These offsets can also be a consequence of substantial obscuration of the AGN optical emission due to the dust surrounding the central SMBH \citep{Hickox2018}. 

In general, offset--AGN are a signature of post-merger galaxies. Interestingly, optical spectroscopic properties of CLASS~B1608+656 reported by \citeauthor{Fassnacht1996} revealed that the optical spectrum of the AGN host galaxy is consistent with ``E+A" or post-starburst galaxies, which are characterised by strong Balmer lines (in absorption) and weak emission lines \citep[e.g.][]{Dressler1983}. Post-starburst galaxies are believed to represent the transition phase from highly star forming galaxies (``blue" galaxies) into ``red" passive galaxies (e.g. \citealt{Pawlik2016} and references therein). Therefore, post-starburst galaxies trace a crucial stage of galaxy evolution.  However, different merging histories can produce this type of galaxy, and characterizing post-merger galaxies can help to disentangle the various evolutionary routes \citep{Pawlik2019}. 
Furthermore, there is growing evidence that galaxy merger may be an important driver for the co-evolution of SMBH and their host galaxies \citep[e.g.][]{Volonteri2007}, and be responsible for the triggering of the AGN activity \citep[e.g.][]{Comerford2015}. 

Confirming that CLASS~B1608+656 is indeed an offset--AGN would provide important insights into a critical stage of galaxy evolution. One way to improve the constraints on the offset would be to apply a pixellated source reconstruction \citep[e.g.][]{Vegetti2009}. This approach would allow us to recover the entire surface brightness of the source and infer also the size and the structure of the AGN host galaxy \citep[e.g.][]{Ritondale2019, Spingola2019sharp}. 
For example, such methods could test whether the optical emission is consistent with a disturbed morphology, which could provide further evidence for a post-merger scenario for CLASS~B1608+656. 
Nevertheless, such sophisticated source reconstructions are computationally expensive, while the parametric approach presented in this work is faster and, therefore, more appropriate for quick initial identification of offset--AGN candidates. Moreover, the fast parametric approach is also very well suited for an automatic search of lensed offset--AGN candidates in the upcoming large-scale surveys.

\subsection{Population of offset--AGN candidates}
\label{sec:PopulationOffsetAGN}

\begin{figure}
\centering
	\includegraphics[width = 0.49\textwidth]{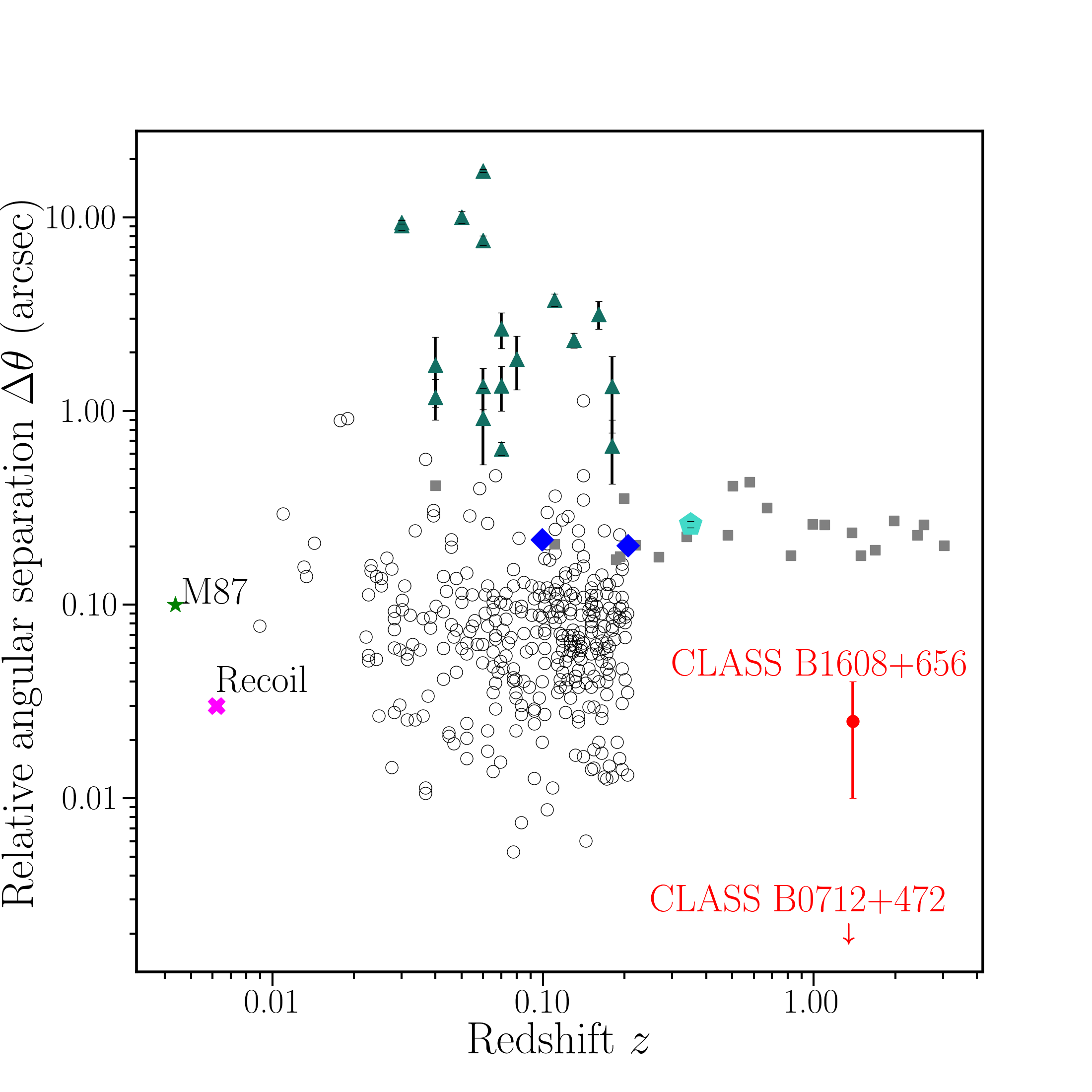}
    \caption{Relative optical--radio (or X-ray) angular separation ($\Delta\theta$ in arcsec) as a function of redshift. The sources plotted here include CLASS~B1608+656 (red) and lower limits for CLASS~B0712+472, and offset--AGN samples found in literature. Black empty circles indicate the sample of \citet{Skipper2018}, which consists of 345 galaxies with offsets estimated using SDSS and VLA observations. Green triangles indicate the offsets from \citet{Barrows2016} estimated using X-ray \textsl{Chandra} imaging and SDSS. The sample of \citet{Orosz2013} obtained using SDSS photometry and VLBI is plotted with grey squares, while blue diamonds are the two dual AGN candidates, J1006+3454 and J1301+4634, in their sample. The green star indicates the offset estimate in M87 from \citet{Batcheldor2010}. Magenta cross shows the average offset and redshift of the recoil SMBH sample from \citet{Lena2014}. The cyan pentagon indicates the offset of the candidate recoil SMBH CXO J101527.2+625911 \citep{Kim2017}. 
    }
    \label{Fig:comparison_offsets}
    \end{figure}

We compare our constraints of the optical--radio offsets in CLASS~B1608+656 and CLASS~B0712+472 with the values for offset--AGN reported in literature. Fig.~\ref{Fig:comparison_offsets} shows this comparison, including confirmed and candidate sources identified using different methods and observations. We include a sample of 345 galaxies from \citeauthor{Skipper2018} (black empty circles), for which the offsets between the radio and optical emissions was estimated using SDSS photometry and VLA A-array observations at 8.4 GHz. The optical--radio offsets in the sample of \citeauthor{Barrows2016} (green triangles) are estimated using X-ray \textsl{Chandra} imaging and SDSS (Table 2 of \citealt{Barrows2016}). We also overlay the two dual AGN candidates (blue diamonds) in the sample of \citeauthor{Orosz2013}, J1006+3454 and J1301+4634, together with their entire sample (grey squares), for which the optical--radio offset is estimated using SDSS photometry and VLBI. We only plot sources from \citet{Orosz2013} with known redshift, and we excluded the optical--radio offsets associated with gravitational lensing systems. Moreover, following their discussion (Sec.~5 of \citealt{Orosz2013}), their sample may be contaminated by AGN with bright radio jets, or gravitational lensing systems with sub-arcsec image separation. We also plot the M87 (green star, \citealt{Batcheldor2010}), the recoil SMBH candidate CXO J101527.2+625911 (cyan pentagon, \citealt{Kim2017}) and the average offset and redshift of the recoil SMBH from \citealt{Lena2014} (magenta cross). In red, we show our results for CLASS~B1608+656 and CLASS~B0712+472 (upper limit). 

\citeauthor{Skipper2018} provide the largest sample with the smallest angular separation between the SDSS and VLA components down to $5\,$mas. However, Fig.~\ref{Fig:comparison_offsets} shows how severely the identification of optical--radio offsets is affected by angular resolution, astrometry, and brightness of multi-wavelength observations.
Moreover, it shows that the optical--radio offset samples available to date probe mainly redshifts lower than 0.2. 
However, to clearly assess the role of galaxy merging in the evolution of active galaxies, it is crucial to significantly expand these samples to redshifts larger than 1. In the sample of \citeauthor{Orosz2013}, there are  sources at high redshift showing optical--radio offsets,  but the offsets are greater than $200\,$mas, while the most dynamic phase of evolution of galaxies, including, for example, interaction of SMBH binaries, happens at much smaller scales (see Sec.~\ref{sec:GalaxyEvolution_LargeSample}).   
CLASS~B1608+656 and CLASS~B0712+472 provide not only an example of high redshift sources with z$\gtrsim 1.3$, but also constrain the optical--radio offsets at scales of dozen of parsecs, which are inaccessible for un-lensed sources, and have the potential to probe even smaller scales. 
The CLASS sample of flat-spectrum gravitationally lensed sources provides observations that can be used to probe a redshift range between 0.6 and 3.6 \citep{Myers2003, Browne2003}, which will be investigated in a later work. Our method applied to the entire CLASS sample will provide a larger statistical sample for searching for offset and multiple AGN systems to constrain SMBH formation and evolution.

%--------------------- Figure offset AGN in other surveys
%--------------------- 

\subsection{Tracing the evolution of active galaxies using large samples of offset--AGN} 
\label{sec:GalaxyEvolution_LargeSample}

The upcoming synoptic all-sky surveys including the Square Kilometer Array (SKA), Euclid, and Legacy Survey of Space and Time (LSST, or  ``Vera C. Rubin" Observatory) will increase the number of gravitationally-lensed AGN from $\sim 10^2$ known today to $\sim 10^5$ \citep{Koopmans2004, Oguri2010, Collett2015, McKean2015}. The method used in this paper can be easily applied to a large sample of gravitationally lensed active galaxies observed at both radio and optical wavelengths. Active galaxies that are gravitationally lensed into four images (i.e caustic configuration as proposed by \citealt{Barnacka2017,Barnacka2018}) can be efficiently identified at both radio and optical wavelengths in current surveys \citep[e.g.][]{Browne2003, Jackson2007}, and it will be even easier with the higher angular resolution of future observing facilities. For example, SKA will provide an angular resolution of $\sim 20\,$mas at 1 GHz and $\sim 2\,$mas at 10 GHz \citep{Dewdney2009,Godfrey2012}, which is an order of magnitude better than the angular resolution of the CLASS survey \citep{Myers2003, Browne2003}. This high angular resolution (if available also for optical observations) should, in principle, allow identification directly based upon the first discovery images of the lensing systems any optical--radio offset at 1--10 mas-level. Moreover, it will provide already stringent constraints to the lens mass model, which is important for reconstructing at high precision the relative position of the radio and optical source components, as demonstrated in this work. Further optical observations of these targets with dedicated observations or matched objects in Gaia or LSST object lists will allow us to constrain the radio optical offsets. 

The identified offset--AGN candidates can then be further investigated with dedicated follow-up at milliarcsecond angular resolution at the radio wavelengths with SKA--VLBI \citep{Paragi2015} and at the optical/near-infrared wavelengths with future adaptive optics instrumentation, such as MICADO at the Extremely Large Telescope (E-ELT, \citealt{Massari2016, Davies2018}).

The observations of gravitationally lensed AGN in caustic configuration have the potential to provide unique insights also into the location of their X-ray emission \citep[e.g.][]{Chartas2019}. The angular resolution of the Chandra Observatory is $\sim$0.5~arcsec. Only the future Lynx\footnote{https://www.lynxobservatory.com} telescope will have comparable resolution at X-rays albeit with much larger effective area. Thus, amplification due to gravitational lensing will be necessary to spatially resolve the X-ray emission at scales much smaller than 0.5~arcsec in high redshift AGN. Once located, the X-ray emission of the lensed source can be compared with the location of the optical emission, and used to find offset-AGN candidates.  

X-ray observations are extremely valuable to search for binary SMBH systems, as they can also unveil highly dust-obscured AGN, which are not necessarily radio-loud \citep[e.g.][]{Vito2019}. The occurrence of SMBH binaries is crucial to our understanding of galaxy formation, as they directly indicate that galaxies evolve through mergers \citep[e.g.][]{1992AJ....104.1039S}. 
Such SMBH binaries are expected to merge, and produce outbursts of gravitational wave emission, whose detection can provide insights into galaxy merger rates and SMBH masses \citep{2019NatAs...3....8M}. 
However, the separation of SMBH binaries emitting gravitational waves must be smaller than 0.1 pc \citep{2018MNRAS.473.3410R}. For this reason, the study of binary (and, in general, multiple) SMBH coalescence is challenging due to the minimum angular resolution required spatially resolve and to monitor such systems \citep{2013ApJ...777...44J}. For example, one of the most compelling observations of SMBH binary is the radio galaxy 0402+379 with two compact-core sources separated by a projected distance of 7.3 pc \citep{2017ApJ...843...14B}. However, 0402+379 is a very nearby source, located at redshift $z=0.055$, which corresponds to luminosity distance of only $\sim 220\,$Mpc. 
Gravitational lensing can provide the necessary spatial resolution to resolve SMBH binary systems like 0402+379 even at high redshift (see also \citealt{Spingola2019c}). We show that it is possible to confine the radio emission of the lensed AGN down to $0.4\,$pc ($0.05\,$mas) for CLASS~B0712+472 and $0.08\,$pc ($9$~$\mu$as) for CLASS~B1608+656. If the systems analyzed here were binary radio-loud AGN, as 0402+379, the lensing amplification would clearly separate the radio emission from the two AGN on parsec-scales.
Thus, gravitational lensing is a promising tool to access the crucial scales needed to provide observational constraints to assess the final parsec problem \citep{2018MNRAS.473.3410R}. 

Finally, large area multi-band surveys with the next-generation of telescopes will allow us to identify and confirm \textsl{un-lensed} offset--AGN candidates, as well as, multiple AGN systems at high redshift \citep{Blecha2018, Burke-Spolaor2018}. As a result, this would provide a parent population to be compared with sample of gravitationally lensed active galaxies showing optical--radio offsets.

\section{Conclusions}
\label{sec:conclusions}

We presented a multi-wavelength analysis of two gravitational lensing systems, CLASS~B0712+472 and CLASS~B1608+656, at redshifts 1.34 and 1.394, respectively, using radio VLBI and optical {\it HST} observations. Our new VLBA observations at 5~mas angular resolution of CLASS~B1608+656 are currently the highest angular resolution radio observations for this system. Both sources are located close to the caustics of their lensing galaxy, and as a result they are highly magnified in a non-linear way.  After correcting for the lensing distortion, we could assess the position of the radio and optical emissions. We estimated the positional uncertainty using Monte Carlo approach that takes into account both the uncertainties related to the observations and the errors of the lens mass model parameters.

We find that in CLASS~B0712+472 the optical and radio emission are co-spatial of $2\pm 5\,$mas, which at redshift of the source, corresponds to an optical--radio offset of $17\pm 42\,$pc.  Conversely, in CLASS~B1608+656 we find an offset between radio emission of $214\pm137\,$ pc. The optical--radio offset in combination with its spectroscopic properties of a post-merger galaxy \citep{Fassnacht1996} allow us to identify CLASS~B1608+656 as a good candidate for a high redshift offset--AGN. If confirmed, CLASS~B1608+656 would be one of a few offset--AGN known at $z>1$, 
and would help constraining the fraction of offset-AGN systems at high redshift \citep[e.g.][]{Volonteri2008}. 

In this work, we showed that gravitational lensing allows us to improve the spatial resolution by at least an order of magnitude for these two systems.
Our multi-wavelength source reconstruction using the state-of-the-art radio and optical facilities and precise lens models allowed us to localize the radio emission down to $0.05\,$mas ($0.4\,$pc) in CLASS~B0712+472, and $0.009$ mas ($0.08\,$pc) in CLASS~B1608+656.  

Further improvement in measuring the optical--radio offset is only limited by the relative astrometry of optical lensed images. Today, the search for offset--AGN is mainly limited to low-redshift galaxies due to the intrinsic limitation of current facilities. However, gravitational lensing eases the identification of offset--AGN at high redshift, in a range that is most critical to test galaxy evolution models.

Enlarging the sample of radio-loud lensed AGN is needed to fully explore the presence of offset--AGN at high redshift. 
Searches for new radio-loud lensing systems in wide-field VLBI surveys are being carried out \citep{Spingola2019mJIVE}, but  in the near future the SKA is expected to find $\sim10^5$ strong lensing systems \citep{Koopmans2004, McKean2015}. If the future radio surveys will have matched sky areas at the optical wavelenghts, it will be possible to discover many gravitationally lensed offset--AGN at high redshift. 

\section*{Acknowledgements}

%people
 We thank the anonymous referee for their constructive comments, which helped clarifying the manuscript. The authors thank Dan Schwartz for his valuable comments and stimulating discussions on this work. We thank Adam Sulkowski for help with proofreading.

%funds
CS is grateful for support from the National Research Council of Science and Technology, Korea (EU$-$16$-$001). This work has been partially supported by NWO-CAS grant 629.001.023.

%data
The results of this paper are based on data obtained from the National Radio Astronomy Observatory's Very Long Baseline Array (VLBA), projects BS251 and BS257. The National Radio Astronomy Observatory is a facility of the National Science Foundation operated under cooperative agreement by Associated Universities, Inc. This paper is based on observations made with the NASA/ESA Hubble Space Telescope, obtained from the data archive at the Space Telescope Science Institute. STScI is operated by the Association of Universities for Research in Astronomy, Inc. under NASA contract NAS 5$-$26555. These observations are associated with programs 9138 and 10158 (see also Table 2).

%software
This research made use of Photutils, an Astropy package for detection and photometry of astronomical sources \citep{Bradley2019}. This research made use of APLpy, an open-source plotting package for Python \citep{Robitaille2012}.

%%%%%%%%%%%%%%%%%%%%%%%%%%%%%%%%%%%%%%%%%%%%%%%%%%

%%%%%%%%%%%%%%%%%%%% REFERENCES %%%%%%%%%%%%%%%%%%

% The best way to enter references is to use BibTeX:
\bibliographystyle{mnras}
\bibliography{paper_references} % if your bibtex file is called example.bib

% Alternatively you could enter them by hand, like this:
% This method is tedious and prone to error if you have lots of references
%\begin{thebibliography}{99}
%\bibitem[\protect\citeauthoryear{Author}{2012}]{Author2012}
%Author A.~N., 2013, Journal of Improbable Astronomy, 1, 1
%\bibitem[\protect\citeauthoryear{Others}{2013}]{Others2013}
%Others S., 2012, Journal of Interesting Stuff, 17, 198
%\end{thebibliography}

%%%%%%%%%%%%%%%%%%%%%%%%%%%%%%%%%%%%%%%%%%%%%%%%%%

%%%%%%%%%%%%%%%%% APPENDICES %%%%%%%%%%%%%%%%%%%%%

\appendix
\section{Robustness of lens mass models}

\begin{figure*}
    \centering
    \includegraphics[width = 0.48\textwidth]{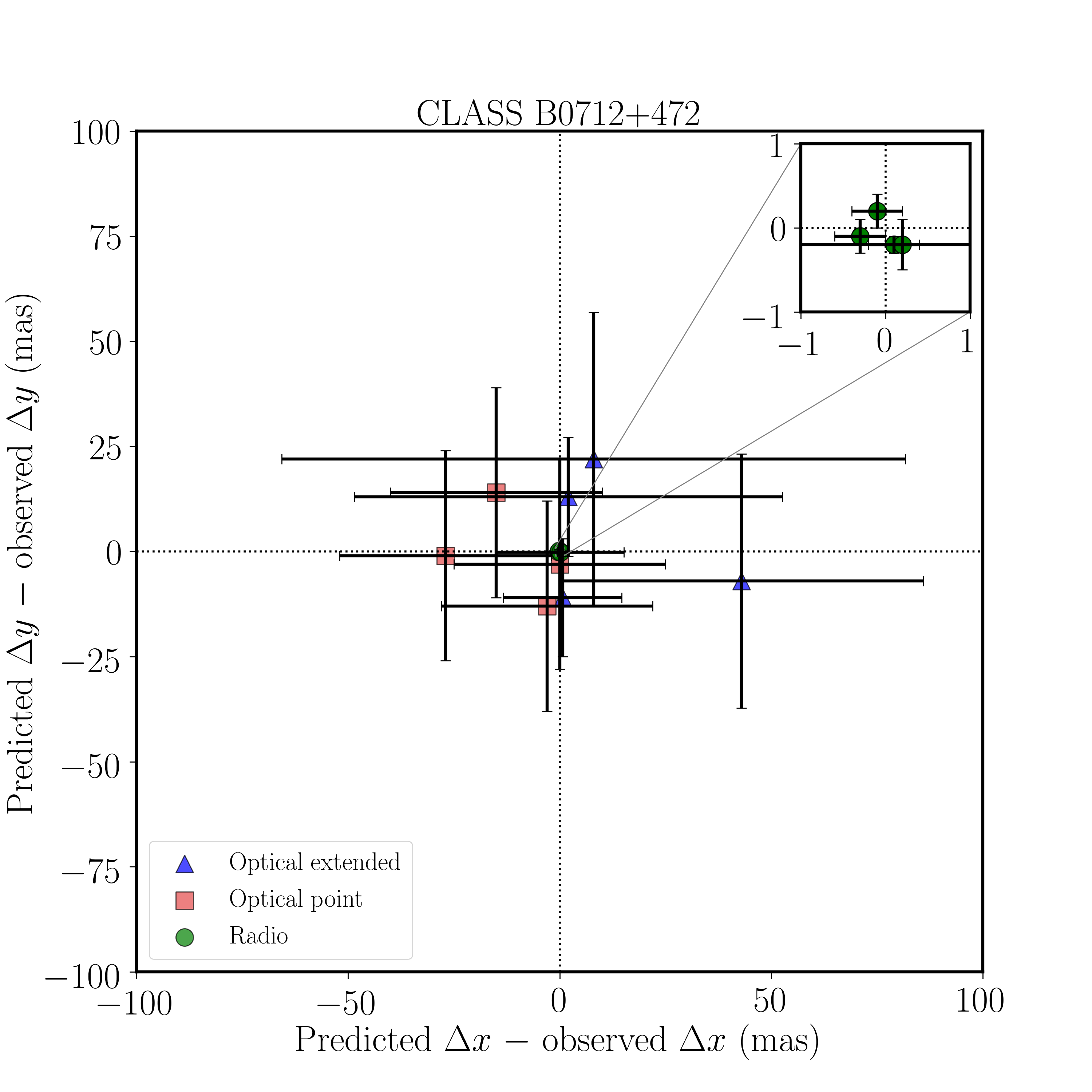}
    \includegraphics[width = 0.48\textwidth]{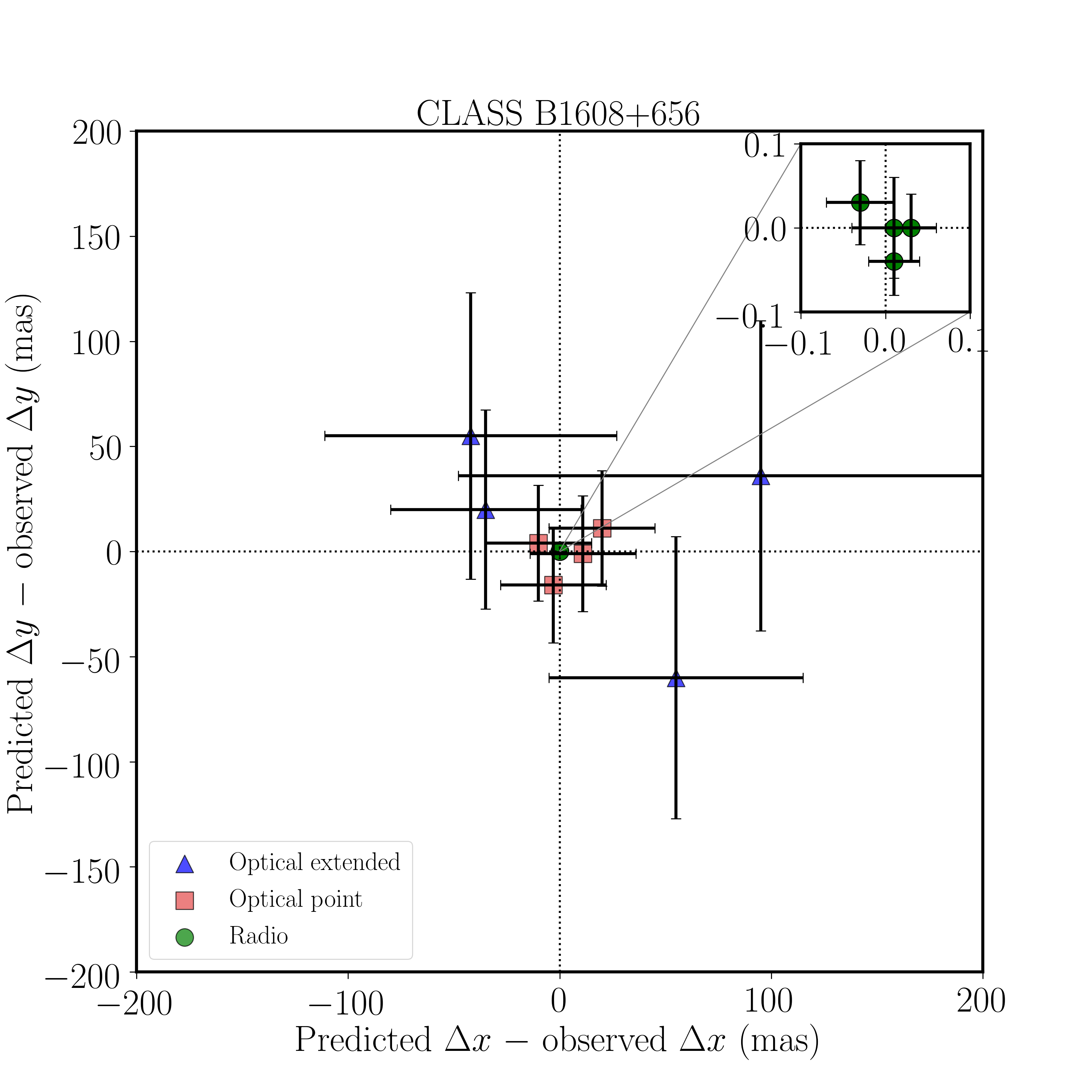}
    \caption{ Offsets between the model-predicted and observed positions of the lensed images (mas) for CLASS~B0712+472 (on the left) and CLASS~B1608+656 (on the right). Each colour and symbol represents a different data set as indicated in the legend on the bottom-left of each plot. The error bars ($1\sigma$ uncertainty) are shown in black and the two black dashed lines indicate the no offset position. The inset on the top-right corner shows a zoom for the measured offsets in the radio, same units.}
    \label{fig:appendix}
\end{figure*}

A lens mass model can be considered robust if it fully recovers the observed properties of the lensed images, in particular their position and flux ratios. In our backward-ray tracing lens modelling approach to reconstruct the multi-wavelength emission in the source plane of CLASS~B0712+472 and CLASS~B1608+656, we adopted the lens models from \citet{Hsueh2017} and \citet{Koopmans2003b}, respectively. 
We did not use the flux ratios, but only the position of the lensed images to recover the centroid of the radio and optical emission in the source plane. Therefore, in order to show the goodness of the mass models, we compare in Fig.~\ref{fig:appendix} the difference between the observed and model-predicted positions of the lensed images. These models can reproduce the observed position of the lensed images within 1$\sigma$, demonstrating that they provide a good representation of the data.

%%%%%%%%%%%%%%%%%%%%%%%%%%%%%%%%%%%%%%%%%%%%%%%%%%

% Don't change these lines
\bsp	% typesetting comment
\label{lastpage}
\end{document}